\documentclass[12pt]{article}
\usepackage{amsmath,amssymb,amscd,amsthm,latexsym}

\font\block=msbm10
\def\C{\hbox{\block\char'0103}}
\def\Z{\hbox{\block\char'0132}}

\font\gotic=eufm10
\def\g{\hbox{\gotic\char'0147}}
\def\l{\hbox{\gotic\char'0154}}
\def\s{\hbox{\gotic\char'0163}}
\def\p{\hbox{\gotic\char'0160}}
\def\o{\hbox{\gotic\char'0166}}

\def\h{\hbox{\gotic\char'0150}}
\def\e{\hbox{\gotic\char'0145}}
\def\i{\hbox{\gotic\char'0151}}

\font\new=eusm10
\def\A{\hbox{\new\char'0101}}

\def\W{\hbox{\new\char'0127}}

\font\yes=msam10
\def\Q{\hbox{\yes\char'0003}}

\begin{document}

\noindent
{\Large Matrix realizations of exceptional superconformal algebras}
\vskip 0.3in
{\large Elena Poletaeva}
\vskip 0.1in


{\it Department of Mathematics,
University of Texas-Pan American,}

{\it Edinburg, TX 78539}

{\it Electronic mail:} elenap$@$utpa.edu

\vskip 0.5in

{\footnotesize \noindent {\bf Abstract.}
We give a general construction of realizations of the contact superconformal algebras $K(2)$ and $\hat{K}'(4)$,
and the exceptional superconformal algebra $CK_6$ as subsuperalgebras of matrices over a Weyl algebra
of size $2^N\times 2^N$, where $N = 1, 2$ and $3$.
We show that there is no such a realization for $K(2N)$, if $N\geq 4$.}
\vskip 0.5in
\noindent
{\it MSC:} 17B65, 17B66, 17B68, 81R10
\vskip 0.1in
\noindent
{\it JGP SC:} Lie superalgebras
\vskip 0.1in
\noindent
{\it Keywords:} Superconformal algebra, Poisson superalgebra, pseudodifferential symbols, Weyl algebra.

\vskip 0.5in
\noindent
{\bf 1. Introduction}
\vskip 0.2in

Superconformal algebras are superextensions of the Virasoro algebra.
They play an important r\^{o}le in the string theory, conformal field theory and mirror symmetry,
and have been extensively studied by mathematicians and physicists.
A {\it superconformal algebra} is a simple complex Lie superalgebra,
spanned by the coefficients of a finite family of
pairwise local fields $a(z) = \sum_{n\in\Z}a_{(n)}z^{-n-1}$,
one of which is the Virasoro field $L(z)$ [3, 8, 9]. It can also be described in terms of vector fields and symbols of differential operators.

An important class of superconformal algebras are the Lie superalgebras $K(N)$ of contact vector fields
on the supercircle $S^{1|N}$ with even coordinate $t$ and $N$ odd coordinates.
The superalgebra $K(N)$ is
characterized by its action on a contact $1$-form [3, 4, 9, 13].
It is spanned by $2^{N}$ fields.
These superalgebras are also known to physicists as the $SO(N)$ superconformal algebras [1, 2].
They are especially interesting, when $N$ is small.
The universal central extension of $K(2)$ is isomorphic to the ``$N = 2$ superconformal algebra''.
The superalgebra $K'(4)$ has three independent central extension, one of which is given by the Virasoro 2-cocycle, and it is isomorphic to
 the ``big $N = 4$ superconformal algebra", see [1, 2].
In this work we consider a different non-trivial central extension $\hat{K}'(4)$ of $K'(4)$.
Note that $K(N)$ has no non-trivial central extensions, if $N > 4$ [13].
The superalgebra $K(6)$ contains the exceptional ``$N = 6$ superconformal algebra'' as a subsuperalgebra.
It constitutes ``one half'' of $K(6)$, and
it is also denoted by $CK_6$, see [3, 6, 9--12,  22--24].

In [17, 18], we proved that for every $N\geq 0$, there exists an embedding of $K(2N)$ into the Poisson superalgebra $P(2N)$
of pseudodifferential symbols on the supercircle $S^{1|N}$.
$P(2N) = P\otimes \Lambda(2N)$, where $P$ is
the Poisson algebra of functions on the cylinder ${T}^*S^1\backslash S^1$,
and $\Lambda(2N)$ is the Grassmann algebra.

It is a  remarkable fact that $K(2)$,
$\hat{K}'(4)$ and $CK_6$, for $N = 1, 2$ and $3$, respectively,
admit embeddings into the family $P_h(2N)$ of Lie superalgebras of
pseudodifferential symbols on $S^{1|N}$, which contracts to $P(2N)$ [17, 18].
Such embeddings allow us to obtain realizations of these superconformal algebras
as subsuperalgebras of matrices of size
$2\times 2$, $4\times 4$, and $8\times 8$, respectively,
over a Weyl algebra $\W = \sum_{i\geq 0}\A d^i,$ where
$\A = \C[t, t^{-1}]$ and $d = {\partial\over \partial t}$, see [19, 20].

In [15] and [16] C. Martinez and E. I. Zelmanov obtained $CK_6$
as a particular case of superalgebras $CK(R, d)$,
where $R$ is an associative commutative superalgebra with an even derivation $d$.
They also realized
$CK_6$ as a subsuperalgebra of matrices of size
$8\times 8$ over $\W$.

In this work we give a general construction of matrix realizations of
$K(2)$, $\hat{K}'(4)$ and $CK_6$.
Note that a semi-direct sum of the Lie algebra
$\o(2N, \C)$ and the Heisenberg Lie superalgebra $\h\e\i(0|2N)$
can be embedded into
the Clifford superalgebra $C(2N)$ and, correspondingly, it has a representation in the Lie superalgebra
$\hbox{End}(\C^{2^{N-1}|2^{N-1}})$, which is related to the spin representation of $\o(2N+1, \C)$ in $\hbox{End}(\C^{2^{N}})$, see [5, 21].
This representation allows to realize the Lie superalgebra $\s\p\o(2|2N)$, which preserves a non-degenerate super skew-symmetric form
on a $(2|2N)$-dimensional superspace,
 as a subsuperalgebra of
$\hbox{End}(\W^{2^{N-1}|2^{N-1}})$.
We prove that if $N = 1, 2$ and $3$, then
$\s\p\o(2|2N)$ and the loop algebra of $\o(2N, \C)$
generate a subsuperalgebra of $\hbox{End}(\W^{2^{N-1}|2^{N-1}})$,
which is isomorphic to
$K(2)$, $\hat{K}'(4)$ and $CK_6$, respectively.
If $N\geq 4$, then the generated superalgebra is the entire
$\hbox{End}(\W^{2^{N-1}|2^{N-1}})$.
Using this fact, we prove that if $N\geq 4$, then there is no embedding of $K(2N)$ into $\hbox{End}(\W^{2^{N-1}|2^{N-1}})$.

In conclusion, we would like to point out that embeddings of
superconformal algebras into Lie superalgebras
of pseudodifferential symbols on a supercircle
and into Lie superalgebras of matrices over a Weyl algebra (which are closely related to each other)
are only possible for superconformal algebras, which are in a sense, exceptional,
and they do not occur in the general case.
This singles out exceptional superconformal algebras from all  superconformal algebras. It would be interesting to give a rigorous
mathematical formulation of this fact.

\vskip 0.2in
\noindent{\bf 2. Preliminaries}
\vskip 0.2in

Let $\Lambda(2N)$ be the Grassmann algebra in $2N$ variables
$\xi_1, \ldots, \xi_N, \eta_1, \ldots, \eta_N $, and
let $\Lambda(1, 2N) =\C [t, t^{-1}]\otimes \Lambda (2N)$ be the associative
superalgebra with natural multiplication and with the following parity
of generators: $p(t) = \bar{0}$, $p(\xi_i) = p(\eta_i) = \bar{1}$
for $i = 1, \ldots, N$. Let
$W(2N)$ be the Lie superalgebra of all superderivations of
$\Lambda(1, 2N)$.
Let $\partial_t$, $\partial_{\xi_i}$ and $\partial_{\eta_i}$
stand for $\partial\over {\partial t}$,
$\partial\over {\partial \xi_i}$ and
$\partial\over {\partial \eta_i}$, respectively.
By definition,
$$
K(2N) = \lbrace D \in W(2N)\mid D\Omega  = f\Omega \hbox{ for some }
f\in \Lambda(1,2 N)\rbrace, \eqno (1)
$$
where
$\Omega = dt + \sum_{i=1}^N \xi_id\eta_i + \eta_id\xi_i$
is a differential 1-form, which is called a {\it contact form}, see [3, 4, 9, 13].
Recall that $K(2N)$ can be described in terms of pseudodifferential symbols
on $S^{1|N}$, see [17, 18].  Consider
the {\it Poisson superalgebra of pseudodifferential symbols}
$$P(2N) = P\otimes \Lambda(2N),\eqno (2)$$
where
the Poisson algebra $P$
is formed by the formal series of the form
$$A(t, \tau) = \sum_{ i = -\infty}^ka_i(t)\tau^i, \eqno (3)$$
where $k$ is some integer,  $a_i(t)\in \C[t, t^{-1}]$, and the even variable $\tau$ corresponds to $\partial_t$, see [14].
The Poisson super bracket is defined as follows:
$$\lbrace A, B\rbrace = \partial_{\tau}A\partial_tB - \partial_tA\partial_{\tau}B +
(-1)^{p(A)+1}\sum_{i = 1}^N(\partial_{\xi_i}A\partial_{\eta_i}B + \partial_{\eta_i}A\partial_{\xi_i}B). \eqno (4)$$
Note that there exists an embedding
$$K(2N)\subset P(2N), \qquad N\geq 0. \eqno (5)$$
Consider a $\Z$-grading of the associative superalgebra
$$P(2N) = \oplus_{i\in\Z}P_{(i)}(2N) \eqno (6)$$
defined by $\hbox{deg}_{Lie}f = \hbox{deg}f - 1$, where
$\hbox{deg}f$ is defined  by
\begin{equation*}
\begin{aligned}
&\hbox{deg } t = \hbox{deg } \eta_i = 0 \hbox{ for } i = 1, \ldots, N, \\
&\hbox{deg } \tau = \hbox{deg } \xi_i = 1 \hbox{ for } i = 1,
 \ldots, N.
 \end{aligned}
 \tag{7}
 \end{equation*}
With respect to the Poisson super bracket,
$$\lbrace P_{(i)}(2N), P_{(j)}(2N)\rbrace \subset P_{({i+j})}(2N).\eqno (8)$$
Thus $P_{(0)}(2N)$ is a subsuperalgebra of $P(2N)$.
We  proved in [17] that $P_{(0)}(2N)$ is isomorphic to $K(2N)$.
Note that $K(2N)$ is spanned by $2^{2N}$ fields, one of which is a Virasoro field.
Recall that a Lie superalgebra is called simple, if it contains no nontrivial ideals [7].
If $N \not= 2$, then
$K(2N)$ is simple. If $N = 2$, then $K(4)$ is not simple. In this case
the derived Lie superalgebra
$K'(4) = [K(4), K(4)]$
 is an ideal in $K(4)$
of codimension one, defined from the exact sequence
$$0\rightarrow K'(4)\rightarrow K(4)\rightarrow \C t^{-1}\tau^{-1}\xi_1\xi_2\eta_1\eta_2\rightarrow 0, \eqno (9)$$
and $K'(4)$ is  simple.
Thus $K'(4)$ is spanned by  16 fields inside $P(4)$. Each field consists of elements, which are indexed by $n$, where $n$ runs through $\Z$. These fields are
\begin{equation*}
\begin{aligned}
&L_n = t^{n+1}\tau, \quad Q_n = t^{n+1}\tau\eta_1\eta_2,\quad
X_n^i = t^{n+1}\tau\eta_i, \\
&Y_n^i = t^{n}\xi_i,\quad
R_n^{ji} = t^{n}\eta_j\xi_i, \quad
Z_n^i = t^{n}\eta_1\eta_2\xi_i,\quad i, j = 1, 2,\\
\end{aligned}
\tag{10}
\end{equation*}
\begin{equation*}
\begin{aligned}
&G_{n}^0 = t^{n-1}\tau^{-1}\xi_1\xi_2,\quad
G_{n}^i = t^{n-1}\tau^{-1}\xi_1\xi_2\eta_i, \quad i = 1, 2,\\
&G_{n}^3 = nt^{n-1}\tau^{-1}\xi_1\xi_2\eta_1\eta_2, \quad n\not=0.
\end{aligned}
\tag{11}
\end{equation*}
Note that $K'(4)$ has three independent central extensions [13].
 $K(6)$ contains the exceptional  superconformal algebra
$CK_6$ as a subsuperalgebra [3, 6,  9--12, 22--24]. $CK_6$ has no non-trivial central extensions [3].
In [18] we obtained a realization of $CK_6$ in terms of pseudodifferential symbols
on $S^{1|3}$, and proved
that $CK_6$ is spanned by  32 fields
inside $K(6) \subset P(6)$. Each field consists of elements indexed by $n$, where $n$ runs through $\Z$.
Explicitly,  $CK_6$ is spanned by the following 20 fields:
\begin{equation*}
\begin{aligned}
&L_n = t^{n+1}\tau, \quad G_n^i = t^{n+1}\tau\eta_i, \hbox{ where  } i = 1, 2, 3, \\
& \tilde{G}_n^i = t^n\xi_i - nt^{n-1}\tau^{-1}\eta_j\xi_i\xi_j, \hbox{ where } i = 1,j = 2
\hbox{ or } i = 2, j = 3 \hbox{ or } i = 3, j = 1,\\
&T_n^{ij} = t^n\eta_i\xi_j - nt^{n-1}\tau^{-1}\eta_k\eta_i\xi_k\xi_j, \hbox{ where } i, j, k\in
\lbrace 1, 2, 3\rbrace \hbox{ and } i\not= j \not= k,\\
&J^{ij}_n = t^{n+1}\tau\eta_i\eta_j, \hbox{ where } 1\leq i < j \leq 3,\\
&\tilde{J}^{ij}_n = t^{n-1}\tau^{-1}\xi_i\xi_j,\hbox{ where }
1\leq i<j\leq 3,\\
&I_n = t^{n+1}\tau\eta_1\eta_2\eta_3,\\
\end{aligned}
\tag{12}
\end{equation*}
and the following $12$ fields, where
$i = 1, j = 2, k = 3$ \hbox{ or } $i = 2, j = 3, k = 1$ \hbox{ or } $i = 3, j = 1, k = 2$:
\begin{equation*}
\begin{aligned}
&T_n^i = -t^n(\eta_j\xi_j + \eta_k\xi_k) + nt^{n-1}\tau^{-1}\eta_j\eta_k\xi_j\xi_k, \\
&S_n^i = -t^n\eta_i(\eta_j\xi_j + \eta_k\xi_k) + nt^{n-1}\tau^{-1}\eta_i\eta_j\eta_k\xi_j\xi_k,\\
&\tilde{S}_n^i = t^{n-1}\tau^{-1}(\eta_j\xi_j - \eta_k\xi_k)\xi_i,\\
&I_n^i = t^{n-1}\tau^{-1}\eta_i\xi_j\xi_k, 
\end{aligned}
\end{equation*}
Note that $L_n$ is a Virasoro field.

\vfil\eject
\noindent
\noindent{\bf 3. Lie superalgebras of matrices over a Weyl algebra}
\vskip 0.2in

By definition, a {\it Weyl algebra} is
$$\W = \sum_{i\geq 0}\A d^i,\eqno (13)$$
 where $\A$ is an associative commutative algebra and
 $d:\A\rightarrow \A$ is a derivation of $\A$,
 with the relations
$$da = d(a) + ad, \quad a\in\A,\eqno (14)$$
see [15, 16].
Set
$$\A = \C[t, t^{-1}], \hbox{ } d = \partial_t. \eqno (15)$$
Let $\hbox{End}(\W^{m|n})$ be the complex Lie superalgebra of matrices of size $(m+n)\times (m+n)$
over  $\W$. Let $\s\p\o(2|2N)$ be a Lie superalgebra, which preserves an even non-degenerate super skew-symmetric form on the $(2|2N)$-dimensional
superspace.

\noindent
{\bf Lemma 3.1.}
For each $N\geq 1$ there exists an embedding
$$\s\p\o(2|2N)\subset \hbox{End}(\W^{2^{N-1}|2^{N-1}}).\eqno (16)$$
{\bf Proof.}
Let $V = \hbox{Span}(\xi_1, \ldots, \xi_N, \eta_1, \ldots, \eta_N)$.
Let $\h\e\i(0|2N)$ be the Heisenberg Lie superalgebra:
$\h\e\i(0|2N)_{\bar{1}} = V$ with the non-degenerate symmetric bilinear form
$(\xi_i, \eta_i) = (\eta_i, \xi_i) = 1$, and $\h\e\i(0|2N)_{\bar{0}} = \C C$,
where $C$ is a central element in $\h\e\i(0|2N)$.
Let $C(2N)$ be the {\it Clifford superalgebra}
with generators $\xi_i, \eta_i$ and relations
$$\xi_i\xi_j = - \xi_j\xi_i,\quad
\eta_i \eta_j = - \eta_j \eta_i,\quad
\eta_i\xi_j = \delta_{i,j}  - \xi_j\eta_i, \quad i, j = 1, \ldots, N.\eqno (17)$$
Let
$$\iota:  \o(2N, \C)  +\hspace{-3.9mm}\supset \h\e\i(0|2N)\rightarrow C(2N),\eqno (18)$$
where $\o(2N, \C)\cong \Lambda^2(V)$,
be an embedding given by
\begin{equation*}
\begin{aligned}
&\iota(\xi_i\xi_j) = \xi_i\xi_j, \quad
\iota(\eta_i\eta_j) = \eta_i\eta_j,\quad
\iota(\xi_i\eta_j) = \xi_i\eta_j, \quad i \not= j,\\
&\iota(\xi_i \eta_i) = \xi_i \eta_i - {1\over 2},\quad
\iota(\xi_i) = \xi_i, \quad \iota(\eta_i) = \eta_i, \quad \iota(C) = 1.
\end{aligned}
\tag{19}
\end{equation*}
Note that $C(2N)\cong \hbox{End}(\C^{2^{N-1}|2^{N-1}})$.
The elements $\xi_i$ act by multiplication on the superspace
$\Lambda(\xi_1, \ldots, \xi_N)$,
and $\eta_i$ acts as ${\partial_{\xi_i}}$. Hence
there exists an embedding
$$\rho: \o(2N, \C)  +\hspace{-3.9mm}\supset \h\e\i(0|2N)\rightarrow \hbox{End}(\C^{2^{N-1}|2^{N-1}}).\eqno (20)$$
Note that if we consider $V$ as an {\it even} vector space,
then formulas (19) define an embedding of $\o(2N + 1, \C)\cong \Lambda^2 (V)\oplus V$ into the {\it Clifford algebra}
$C(2N)$, and correspondingly, the spin representation
of $\o(2N + 1, \C)$ in $\hbox{End}(\C^{2^{N}})$, see [5, 21].

\noindent
Let
$$\hbox{End}(\C^{2^{N-1}|2^{N-1}}) =
\hbox{End}_{-1}\oplus \hbox{End}_0 \oplus \hbox{End}_1,\eqno (21)$$
where $\hbox{End}_0$ is the set of  even complex matrices, and $\hbox{End}_{-1}$ and $\hbox{End}_{1}$ are the sets
of odd upper triangular matrices and odd lower triangular matrices, respectively.
Let $X \in V$. Then $\rho(X) = \rho(X)_{-1} + \rho(X)_1$, where
$\rho(X)_{\pm 1}\in \hbox{End}_{\pm 1}$.
Define
$$\rho(X)^{\pm} \in \hbox{End}(\W^{2^{N-1}|2^{N-1}}) \eqno (22)$$
by setting
$$\rho(X)^{\pm} = \rho(X)_{-1}^{\pm} + \rho(X)_{1}^{\pm}, \eqno (23)$$
where
\begin{equation*}
\begin{aligned}
&\rho(X)_{-1}^{\pm} = t^{\pm 1}\rho(X)_{-1}, \\
&\rho(X)_{1}^{\pm} = (tdt^{\pm 1} \mp {1\over 2}t^{\pm 1})\rho(X)_{1}.
\end{aligned}
\tag{24}
\end{equation*}
Define  $\g = \g_{\bar 0} \oplus \g_{\bar 1}$ by setting
\begin{equation*}
\begin{aligned}
&\g_{\bar{1}} = \rho(V)^{\pm},\\
&\g_{\bar{0}} = \rho( \o(2N, \C)) \oplus \s\l(2),
\end{aligned}
\tag{25}
\end{equation*}
where
$\s\l(2) = \hbox{Span}(E, H, F),$ and
$E, H$ and $F$ are the following diagonal matrices in  $\hbox{End}(\W^{2^{N-1}|2^{N-1}})$:
\begin{equation*}
\begin{aligned}
&E = {1\over 2}i\Big((tdt^2 - {1\over 2}t^2){1}_{2^{N-1}} \hbox{ } | \hbox{ } (t^2dt - {1\over 2}t^2) 1_{2^{N-1}}\Big),\\
&F = {1\over 2}i\Big((tdt^{-2} + {1\over 2}t^{-2})1_{2^{N-1}} \hbox{ } | \hbox{ }
(dt^{-1} + {1\over 2}t^{-2})1_{2^{N-1}}\Big),\\
&H = (td)1_{2^{N-1}|2^{N-1}},
\end{aligned}
\tag{26}
\end{equation*}
so that the standard commutation relations hold:
$$[H, E] = 2E, \quad [H, F] = -2F, \quad [E, F] = H.$$
Then $\g \cong \s\p\o(2|2N).$
$$\eqno\Q$$
\vskip 0.1in
\noindent
Let $\tilde{\o}(2N, \C) = \hbox{Span}(t^n\rho(X) \hbox{ } | \hbox{ } X\in {\o}(2N, \C), \hbox{ }
n\in \Z)$. Thus $\tilde{\o}(2N, \C)$ is isomorphic to the loop algebra of
$\o(2N, \C)$.
\vskip 0.1in
\noindent
{\bf Theorem 3.2.}
If $N = 1$, then $\s\p\o (2|2)$ and $\tilde{\o}(2, \C)$ generate  $K(2)$,

\noindent
if $N = 2$, then $\s\p\o (2|4)$ and $\tilde{\o}(4, \C)$ generate $\hat{K}'(4)$,

\noindent
if $N = 3$,  then $\s\p\o (2|6)$ and $\tilde{\o}(6, \C)$ generate $CK_6$,

\noindent
if $N\geq 4$,  then $\s\p\o (2|2N)$ and $\tilde{\o}(2N, \C)$ generate
$\hbox{End}(\W^{2^{N-1}|2^{N-1}})$.

\vskip 0.1in
\noindent
{\bf Proof.}
{\it Case $N = 1$.}
It is easy to see that $\s\p\o (2|2)_{\bar{1}} =
 \hbox{Span}(\rho(\xi_1^{\pm}), \rho(\eta_1^{\pm}))$, where
\begin{equation*}
\begin{aligned}
&\rho(\xi_1^+) =
\left(\begin{array}{c|c}
0&0\\
\hline
t&0\\
\end{array}\right), \hbox{ }
\rho(\xi_1^-) =
\left(\begin{array}{c|c}
0&0\\
\hline
t^{-1}&0\\
\end{array}\right),\\
&\rho(\eta_1^+) =
\left(\begin{array}{c|c}
0&t^2d+{1\over 2}t\\
\hline
0&0\\
\end{array}\right), \hbox{ }
\rho(\eta_1^-) =
\left(\begin{array}{c|c}
0&d - {1\over 2}t^{-1}\\
\hline
0&0\\
\end{array}\right).
\end{aligned}
\tag{27}
\end{equation*}
Hence
$\s\p\o (2|2)$ and $\tilde{\o}(2, \C) = \hbox{Span}(
\left(\begin{array}{c|c}
-t^n&0\\
\hline
0&t^n\\
\end{array}\right))$
 generate the following subsuperalgebra  of  $\hbox{End}(\W^{1|1})$:
$$\g =  \hbox{Span} (L_n, H_n, {G}_n,\tilde{G}_n \hbox{ }|\hbox{ } n \in \Z ),$$
where
\begin{equation*}
\begin{aligned}
&L_n = \left(\begin{array}{c|c}
t^{n+1}d + nt^n&0\\
\hline
0&t^{n+1}d\\
\end{array}\right), \quad
H_n = \left(\begin{array}{c|c}
-t^n&0\\
\hline
0&t^n\\
\end{array}\right), \\
&G_n = \left(\begin{array}{c|c}
0&t^{n+1}d + {{n}\over 2}t^{n}\\
\hline
0&0\\
\end{array}\right), \quad
\tilde{G}_n =\left(\begin{array}{c|c}
0&0\\
\hline
t^n&0\\
\end{array}\right).
\end{aligned}
\tag{28}
\end{equation*}
The isomorphism
$$\sigma: K(2)\subset P(2)\rightarrow \g$$
is given by
\begin{equation*}
\begin{aligned}
&\sigma(t^{n+1}\tau) = L_n+ {n\over 2}H_n, \quad
\sigma(t^{n}\xi_1\eta_1) = {1\over 2}H_n, \\
&\sigma(t^{n}\xi_1) = \tilde{G}_n, \quad
\sigma(t^{n+1}\tau\eta_1) = {G}_n.
\end{aligned}
\tag{29}
\end{equation*}
Note that $L_n$ is a Virasoro field.

In the cases when $N = 2$ and $3$, we will use an embedding of
$\hat{K}'(4)$ and $CK_6$, respectively, into a deformation of $P(2N)$.
Let $P_1(2N) = P_1 \otimes C(2N)$.
The associative multiplication in the vector space $P_1 = P$
is determined as follows (see [14]):
$$
A(t, \tau)\circ B(t, \tau) =
\sum_{n\geq 0} {1\over {n!}}\partial^n_{\tau}A(t, \tau)
\partial^n_{t}B(t, \tau).
\eqno (30)$$
 The product of
$A = A_1 \otimes X$ and $B = B_1 \otimes  Y$,
where $A_1, B_1 \in P_1$, and $X, Y \in C(2N)$,
is given by
$$AB =
(A_1\circ B_1)\otimes (X Y). \eqno (31)$$
The Lie super bracket in $P_1(2N)$ is
$[A, B] = AB - (-1)^{p(A)p(B)}BA$.
$P_1(2N)$ is called the  {\it Lie superalgebra of pseudodifferential symbols on $S^{1|N}$}, see [17, 18].

\noindent
{\it Case $N = 2$}.
We proved in [17] that $\hat{K}'(4)$ is spanned inside
$P_1(4)$ by the 12 fields given in (10) and 4 fields
\begin{equation*}
\begin{aligned}
&G_{n}^0 = \tau^{-1}\circ t^{n-1}\xi_1\xi_2,\quad
G_{n}^i = \tau^{-1}\circ t^{n-1}\xi_1\xi_2\eta_i, \quad i = 1, 2,\\
&G_{n}^3 = n\tau^{-1}\circ t^{n-1}\xi_1\xi_2\eta_1\eta_2 + t^n.
\end{aligned}
\tag{32}
\end{equation*}
Note that $L_n$ is a Virasoro field. The central element in $\hat{K}'(4)$ is $G^3_{0} = 1$, and the corresponding $2$-cocycle is
\begin{equation*}
\begin{aligned}
&c(L_n, G_k^3) = -n\delta_{n+k,0},\\
&c(X_n^i, G_k^j) = (-1)^j\delta_{n+k,0}, \quad 1\leq i\not= j\leq 2,\\
&c(Q_n, G_k^0) = \delta_{n+k,0}.
\end{aligned}
\tag{33}
\end{equation*}
Note that this 2-cocycle is different from the Virasoro 2-cocycle.
Let $V^{\mu} = t^{\mu}\C[t, t^{-1}]\otimes\Lambda(\xi_1, \xi_2)$, where $\mu\in\C\backslash\Z$.
Define
a representation of $\hat{K}'(4)$ in $V^{\mu}$ accordingly to the formulas (10) and (32).
In particular, $\tau^{-1}$ is identified with an antiderivative,
and the central element  acts by the identity operator.
Consider the following basis in $V^{\mu}$:
\begin{equation*}
\begin{aligned}
&v_m^0(\mu) = t^{m+\mu},\quad v_m^i(\mu) = t^{m+\mu}\xi_i, \quad i = 1, 2, \\
&v_m^3(\mu) = {t^{m+\mu}\over{m+\mu}}\xi_1\xi_2, \hbox{ }m\in\Z.
\end{aligned}
\tag{34}
\end{equation*}
Explicitly, the action of $\hat{K}'(4)$ on $V^{\mu}$ is given as follows:
\begin{equation*}
\begin{aligned}
&L_n(v_m^i(\mu)) = (m + \mu) v_{m+n}^i(\mu), \hbox{ } i = 0, 1, 2, \\
&L_n(v_m^3(\mu)) = (n+ m + \mu) v_{m+n}^3(\mu),\\
&X_n^i(v_m^i(\mu)) = (m+\mu)v_{m+n}^0(\mu),\hbox{ } i = 1, 2, \\
&X_n^1(v_m^3(\mu)) = v_{m+n}^2(\mu), \\
&X_n^2(v_m^3(\mu)) = - v_{m+n}^1(\mu), \\
&Q_n(v_m^3(\mu)) = -v_{m+n}^0(\mu), \\
&Y_n^i(v_m^0(\mu)) = v_{m+n}^i(\mu),\hbox{ } i = 1, 2,\\
&Y_n^1(v_m^2(\mu)) = (n + m +\mu)v_{m+n}^3(\mu),\\
&Y_n^2(v_m^1(\mu)) =-(n + m +\mu)v_{m+n}^3(\mu),\\
&R_n^{ii}(v_m^0(\mu)) = v_{m+n}^0(\mu),\hbox{ }i = 1, 2,\\
&R_n^{ii}(v_m^j(\mu)) = v_{m+n}^j(\mu),\quad
R_n^{ij}(v_m^i(\mu)) = -v_{m+n}^j(\mu),\hbox{ }i \not= j = 1, 2,\\
&Z_n^1(v_m^2(\mu)) = -v_{m+n}^0(\mu),\quad
Z_n^2(v_m^1(\mu)) = v_{m+n}^0(\mu),\\
&G_{n}^0(v_m^0(\mu)) = v_{m+n}^3(\mu),\quad
G_{n}^i(v_m^i(\mu)) = v_{m+n}^3(\mu),\quad i = 1, 2,\\
&G_{n}^3(v_m^i(\mu)) = v_{m+n}^i(\mu), \hbox{ }n\not= 0, \hbox{ } i = 0, 1, 2, 3.
\end{aligned}
\tag{35}
\end{equation*}
These formulas remain valid for  $\mu = 0$. Thus we obtain a representation of
$\hat{K}'(4)$ in the superspace $\C[t, t^{-1}]\otimes \Lambda(\xi_1, \xi_2)$ with a basis
$$\lbrace v_{m}^0,  v_{m}^3; v_{m}^1,  v_{m}^2 \rbrace,$$ where
$$v_m^0 = t^m,  \quad v_m^3 = t^m\xi_1\xi_2,\quad
v_m^i = t^m\xi_i, \quad i = 1, 2, \quad m\in\Z.$$
We have
\begin{equation*}
\begin{aligned}
&L_n(v_m^i) = t^{n+1}dv_{m}^i, \hbox{ } i = 0, 1, 2,\quad
L_n(v_m^3) = tdt^{n}v_{m}^3\\
&X_n^i(v_m^i) =  t^{n+1}dv_{m}^0,\hbox{ } i = 1, 2, \quad
X_n^1(v_m^3) = t^{n}v_{m}^2, \\
&X_n^2(v_m^3) = - t^{n}v_{m}^1, \quad
Q_n(v_m^3) = -t^nv_{m}^0, \\
&Y_n^i(v_m^0) = t^nv_{m}^i,\hbox{ } i = 1, 2,\quad
Y_n^1(v_m^2) = tdt^nv_{m}^3,\\
&Y_n^2(v_m^1) =- tdt^nv_{m}^3,\quad
R_n^{ii}(v_m^0) = t^nv_{m}^0, \hbox{ }i = 1, 2,\\
&R_n^{ii}(v_m^j) = t^nv_{m}^j, \quad
R_n^{ij}(v_m^i) = -t^nv_{m}^j, \quad i\not= j = 1, 2,\\
&Z_n^1(v_m^2) = -t^nv_{m}^0,\quad
Z_n^2(v_m^1) = t^nv_{m}^0,\\
&G_{n}^0(v_m^0) = t^nv_{m}^3,\quad
G_{n}^i(v_m^i) = t^nv_{m}^3,\quad i = 1, 2,\\
&G_{n}^3(v_m^i) = t^nv_{m}^i, \hbox{ }n\not= 0,\hbox{ }i = 0, 1, 2, 3.
\end{aligned}
\tag{36}
\end{equation*}
This gives a realization of $\hat{K}'(4)$ as a subsuperalgebra of  End($\W^{2|2}$).
Note that
$$\s\p\o(2|4)\subset \hat{K}'(4)\subset \hbox{End}(\W^{2|2}),\eqno (37)$$
where $\s\p\o(2|4)_{\bar{1}}$ is spanned by the following elements:
\begin{equation*}
\begin{aligned}
&\rho(\xi_1)^{\pm} = Y_{\pm 1}^1 \mp {1\over 2}G_{\pm 1}^2, \quad
\rho(\xi_2)^{\pm} = Y_{\pm 1}^2 \pm {1\over 2}G_{\pm 1}^1, \\
&\rho(\eta_1)^{\pm} = X_{\pm 1}^1 \pm {1\over 2}Z_{\pm 1}^2, \quad
\rho(\eta_2)^{\pm} = X_{\pm 1}^2 \mp {1\over 2}Z_{\pm 1}^1.
\end{aligned}
\tag{38}
\end{equation*}
$\tilde{\o}(4, \C)$ is generated by $R_n^{12}, R_n^{21}, G_n^0$ and $Q_n$.
When these elements act on $\s\p\o(2|4)_{\bar 1}$, they generate the 8 fields
$X_n^i,  Y_n^i, G_n^i$ and $Z_n^i$, where $i = 1, 2$,
which span
$\hat{K}'(4)_{\bar 1}$, and hence  $\hat{K}'(4)$ is generated by $\s\p\o(2|4)$ and $\tilde{\o}(4, \C)$.

\noindent
{\it Case $N = 3$.}
We proved in [18] that $CK_6$ is spanned inside $P_1(6)$ by the 8 fields: $L_n$, $G_n^i$, $I_n$, and $J^{ij}_n$, given in (12),
and the following 24 fields, where $n\in\Z$:

\noindent
12 fields:
\begin{equation*}
\begin{aligned}
&\tilde{G}_{n}^i = t^n\xi_i - n\tau^{-1}\circ t^{n-1}\xi_i\xi_j\eta_j, 
\hbox{ where } i = 1,j = 2 \hbox{ or } i = 2, j = 3 \hbox{ or } i = 3, j = 1,\\
&T_{n}^{ij} = t^n\eta_i\xi_j - n\tau^{-1}\circ t^{n-1}\xi_k\xi_j\eta_k\eta_i, \hbox{ where } i, j, k \in \lbrace 1, 2, 3\rbrace
\hbox{ and } i\not= j \not= k,\\
&\tilde{J}_{n}^{ij} = \tau^{-1}\circ t^{n-1}\xi_i\xi_j,
\hbox{ where }  1\leq i<j\leq 3,
\end{aligned}
\tag{39}
\end{equation*}
and the following 12 fields, where
$i = 1, j = 2, k = 3$ \hbox{ or } $i = 2, j = 3, k = 1$ \hbox{ or } $i = 3, j = 1, k = 2$:
\begin{equation*}
\begin{aligned}
&T_{n}^i =  -t^n(\eta_j\xi_j + \eta_k\xi_k) + n\tau^{-1}\circ t^{n-1}\xi_j\xi_k\eta_j\eta_k + t^n,\\
&S_{n}^i = -t^n\eta_i(\eta_j\xi_j + \eta_k\xi_k) + n\tau^{-1}\circ t^{n-1}\xi_j\xi_k\eta_i\eta_j\eta_k
+ t^n\eta_i,\\
&\tilde{S}_{n}^i = \tau^{-1}\circ t^{n-1}(\xi_j\xi_i\eta_j - \xi_k\xi_i\eta_k),\\
&I_{n}^i = \tau^{-1}\circ t^{n-1}\xi_j\xi_k\eta_i.
\end{aligned}
\tag{40}
\end{equation*}

\noindent
Let $V^{\mu} = t^{\mu}\C[t, t^{-1}]\otimes\Lambda(\xi_1, \xi_2, \xi_3)$, where $\mu\in\C\backslash\Z$.
Define a representation of $CK_6$ in $V^{\mu}$ according to the formulas (12) and (39--40).
Consider the following basis in $V^{\mu}$:
\begin{equation*}
\begin{aligned}
&v_m^i(\mu) = {t^{m+\mu}}\xi_i,\quad
\hat{v}_m^i(\mu) = {t^{m+\mu}\over {m+\mu}}\xi_j\xi_k, \hbox{ } 1\leq i\leq 3,\\
&v_m^4(\mu) = t^{m+\mu}, \quad \hat{v}_m^4(\mu) =  -{t^{m+\mu}\over {m+\mu}}\xi_1\xi_2\xi_3,
\end{aligned}
\end{equation*}
where $m \in \Z$ and $(i, j, k)$ is the cycle $(1, 2, 3)$ in the formulas for $\hat{v}^i_m(\mu)$.
Explicitly, the action of $CK_6$ on $V^{\mu}$ is given as follows:
\bigskip
\begin{equation*}
\begin{aligned}
&L_{n}(v_m^i(\mu)) = (m + \mu)v_{m+n}^i(\mu),\quad
L_{n}(\hat{v}_m^i(\mu)) = (m + n + \mu)\hat{v}_{m+n}^i(\mu),\\
&G_{n}^i(v_m^i(\mu)) = (m + \mu)v_{m+n}^4(\mu),\quad
G_{n}^i(\hat{v}_m^4(\mu)) = -(m + n + \mu)\hat{v}_{m+n}^i(\mu),\\
&G_{n}^i(\hat{v}_m^k(\mu)) = {v}_{m+n}^j(\mu), \hbox{ } G_{n}^i(\hat{v}_m^j(\mu)) = -{v}_{m+n}^k(\mu),\\
&\tilde{G}_{n}^i(v_m^4(\mu)) = v_{m+n}^i(\mu), \quad
 \tilde{G}_{n}^i(\hat{v}_m^i(\mu)) = -\hat{v}_{m+n}^4(\mu),\\
&\tilde{G}_{n}^i({v}_m^k(\mu)) = -(m + n + \mu)\hat{v}_{m+n}^j(\mu), \quad
\tilde{G}_{n}^j({v}_m^j(\mu)) = (m  +\mu)v_{m+n}^k(\mu),\\
&T_{n}^{ij}(v_m^i(\mu)) = - v_{m+n}^j(\mu), \quad
T_{n}^{ij}(\hat{v}_m^j(\mu)) =  \hat{v}_{m+n}^i(\mu),\\
&T_{n}^i(v_m^i(\mu)) = - v_{m+n}^i(\mu), \quad
T_{n}^i(v_m^4(\mu)) = - v_{m+n}^4(\mu),\\
&T_{n}^i(\hat{v}_m^i(\mu)) = \hat{v}_{m+n}^i(\mu), \quad
T_{n}^i(\hat{v}_m^4(\mu)) = \hat{v}_{m+n}^4(\mu),\\
&S_{n}^i(v_m^i(\mu)) = - v_{m+n}^4(\mu),\quad
S_{n}^i(\hat{v}_m^4(\mu)) = - \hat{v}_{m+n}^i(\mu),\\
&\tilde{S}_{n}^i({v}_m^k(\mu)) = - \hat{v}_{m+n}^j(\mu), \quad
\tilde{S}_{n}^i({v}_m^j(\mu)) = - \hat{v}_{m+n}^k(\mu),\\
&I_{n}^i({v}_m^i(\mu)) = \hat{v}_{m+n}^i(\mu), \quad
I_{n}(\hat{v}_m^4(\mu)) = {v}_{m+n}^4(\mu),\\
&J_{n}^{ij}(\hat{v}_m^k(\mu)) = - {v}_{m+n}^4(\mu), \quad
J_{n}^{ij}(\hat{v}_m^4(\mu)) = {v}_{m+n}^k(\mu),\\
&\tilde{J}_{n}^{ij}({v}_m^4(\mu)) = \hat{v}_{m+n}^k(\mu), \quad
\tilde{J}_{n}^{ij}({v}_m^k(\mu)) = - \hat{v}_{m+n}^4(\mu),
\end{aligned}
\tag{41}
\end{equation*}
where $(i, j, k)$ is the cycle $(1, 2, 3)$.
These formulas remain valid for $\mu = 0$.
 Thus we obtain a representation of
$CK_6$ in the superspace $\C[t, t^{-1}]\otimes \Lambda(\xi_1, \xi_2, \xi_3)$
with a basis
$$\lbrace \hat{v}_{m}^i, v_{m}^4;
 v_{m}^i,\hat{v}_{m}^4\rbrace,\quad i = 1, 2, 3,$$
  where
$$\hat{v}_m^i = t^m\xi_j\xi_k, \quad
v^4 = t^m, \quad v_m^i = t^m\xi_i, \quad\hat{v}_m^4 = -t^m\xi_1\xi_2\xi_3,$$
\noindent
$(i, j, k)$ is the cycle $(1, 2, 3)$ in the formulas for $\hat{v}_m^i$, and $m\in\Z$.
We have
\begin{equation*}
\begin{aligned}
&L_{n}(v_m^i) = t^{n+1}dv_{m}^i,\quad
L_{n}(\hat{v}_m^i) = tdt^n\hat{v}_{m}^i,\\
&G_{n}^i(v_m^i) = t^{n+1}dv_{m}^4,\quad
G_{n}^i(\hat{v}_m^4) = -tdt^n\hat{v}_{m}^i,\\
&G_{n}^i(\hat{v}_m^k) = t^n{v}_{m}^j, \quad G_{n}^i(\hat{v}_m^j) = -t^n{v}_{m}^k,\\
&\tilde{G}_{n}^i(v_m^4) = t^nv_{m}^i, \quad \tilde{G}_{n}^i(\hat{v}_m^i) = -t^n\hat{v}_{m}^4,\\
&\tilde{G}_{n}^i({v}_m^k) = -tdt^n\hat{v}_{m}^j, \quad
\tilde{G}_{n}^j({v}_m^j) = t^{n+1}dv_{m}^k,\\
&T_{n}^{ij}(v_m^i) = - t^nv_{m}^j, \quad
T_{n}^{ij}(\hat{v}_m^j) =  t^n\hat{v}_{m}^i,\\
&T_{n}^i(v_m^i) = - t^nv_{m}^i, \quad
T_{n}^i(v_m^4) = - t^nv_{m}^4,\\
&T_{n}^i(\hat{v}_m^i) = t^n\hat{v}_{m}^i, \quad
T_{n}^i(\hat{v}_m^4) = t^n\hat{v}_{m}^4, \\
&S_{n}^i(v_m^i) = - t^nv_{m}^4,\quad
S_{n}^i(\hat{v}_m^4) = - t^n\hat{v}_{m}^i,\\
\end{aligned}
\tag{42}
\end{equation*}
\begin{equation*}
\begin{aligned}
&\tilde{S}_{n}^i({v}_m^k) = - t^n\hat{v}_{m}^j, \quad
\tilde{S}_{n}^i({v}_m^j) = - t^n\hat{v}_{m}^k,\\
&I_{n}^i({v}_m^i) = t^n\hat{v}_{m}^i, \quad
I_{n}(\hat{v}_m^4) = t^n{v}_{m}^4,\\
&J_{n}^{ij}(\hat{v}_m^k) = - t^n{v}_{m}^4, \quad
J_{n}^{ij}(\hat{v}_m^4) = t^n{v}_{m}^k,\\
&\tilde{J}_{n}^{ij}({v}_m^4) = t^n\hat{v}_{m}^k, \quad
\tilde{J}_{n}^{ij}({v}_m^k) = - t^n\hat{v}_{m}^4,
\end{aligned}
\end{equation*}
where $(i, j, k)$ is the cycle $(1, 2, 3)$.
This gives a realization of $CK_6$  as subsuperalgebra of End($\W^{4|4}$).
Note that
$$\s\p\o(2|6)\subset CK_6\subset \hbox{End}(\W^{4|4}),\eqno (43)$$
where $\s\p\o(2|6)_{\bar{1}}$ is spanned by the following elements:
\begin{equation*}
\begin{aligned}
&\rho(\xi_i)^{\pm} = \tilde{G}_{\pm 1}^i \mp {1\over 2}\tilde{S}_{\pm 1}^i, \\
&\rho(\eta_i)^{\pm} = {G}_{\pm 1}^i \mp {1\over 2}{S}_{\pm 1}^i,\quad
i = 1, 2, 3.
\end{aligned}
\tag{44}
\end{equation*}
The Lie algebra $\tilde{\o}(6, \C)$ is generated by $T_n^{ij}$ $(i\not= j)$, and $J_n^{ij}, \tilde{J}_n^{ij}$ $i < j$.
Clearly, when these elements act on $\s\p\o(2|6)_{\bar 1}$, they generate the 12 fields
$G_{n}^i,  \tilde{G}_{n}^i,  S_{n}^i$, and  $\tilde{S}_{n}^i$, where $i = 1, 2, 3$.
They also generate the 4 fields: $I_{n}^i$, $i = 1, 2, 3$ and $I_{n}$, due to the
commutation relations
\begin{equation*}
\begin{aligned}
&[\tilde{J}_{n}^{ij}, \rho(\eta_k)^{+}] = -nI_{n+1}^k, \\
&[{J}_{n}^{ij}, \rho(\eta_k)^{+}] = -nI_{n+1},
\end{aligned}
\tag{45}
\end{equation*}
where $(i, j, k)$ is the cycle $(1, 2, 3)$, and ${J}_{n}^{ij} =  - {J}_{n}^{ji}$, $\tilde{J}_{n}^{ij} =  - \tilde{J}_{n}^{ji}$
for $i > j$.
Thus they generate the 16 fields, which span
${(CK_6)}_{\bar 1}$, and hence  $CK_6$ is generated by $\s\p\o(2|6)$ and $\tilde{\o}(6, \C)$.

\noindent
{\it Case $N = 4$.} Let $S$ be the subset of
$\hbox{End}(\W^{2^{N-1}|2^{N-1}})$, generated by $\s\p\o(2|8)$ and $\tilde{\o}(8, \C)$.
Consider the following basis in $\Lambda(\xi_1, \ldots, \xi_4)$:
\begin{equation*}
\begin{aligned}
&\Lambda(\xi_1, \ldots, \xi_4)_{\bar{0}} =
\lbrace v_0, v_{12}, v_{13},v_{14},v_{23},v_{24}, v_{34},\hat{v}_{0}\rbrace,\\
&\Lambda(\xi_1, \ldots, \xi_4)_{\bar{1}} =
\lbrace v_1, \ldots, v_4, \hat{v}_1, \ldots, \hat{v}_4\rbrace,
\end{aligned}
\tag{46}
\end{equation*}
where
$$v_0 = 1, \quad v_{ij} = \xi_i\xi_j, \quad \hat{v}_{0} = \xi_1\xi_2\xi_3\xi_4,
\quad {v}_{i} = \xi_i, \quad
\hat{v}_{i} = \xi_1 \cdots \hat{\xi}_i \cdots\xi_4.$$
Let $E_{\pm 1}^{i,j}$  be an elementary $8\times 8$  matrix
 in $\hbox{End}_{\pm 1}$. Let
\begin{equation*}
\begin{aligned}
&E_1 = \lbrace t^nX, (t^nd)X \hbox{ }| \hbox{ } X = E_{1}^{i,j}, n\in \Z\rbrace,\\
&E_{-1} = \lbrace t^nX \hbox{ }| \hbox{ } X = E_{-1}^{i,j}, n\in \Z\rbrace.
\end{aligned}
\tag{47}
\end{equation*}
Note that it suffices to show that
$E_{\pm 1}\subset S.$
Let $E_{0}^{i,j}$ and $\tilde{E}_{0}^{i,j} = E_{0}^{i+8,j+8}$, where
$1\leq i, j\leq 8$,
be elementary $8\times 8$ matrices in $\hbox{End}_{0}$. Let
$$E_{0} = \lbrace t^nX \hbox{ }| \hbox{ } X = E_{0}^{i,j}, \tilde{E}_{0}^{i,j}, \quad i \not= j, \quad n \in \Z\rbrace.\eqno (48)$$
Note that $E_0\subset S$.
In fact,
$$[t^n\rho(\eta_1\eta_2), \rho(\eta_3)^+] = nt^{n+1}E_1^{1,8},\eqno (49)$$
Similarly, we can show that
$t^n{E}_{1}^{i,j}\in S$, for
\begin{equation*}
\begin{aligned}
&i= 1, j = 5, 6, 7, 8,\quad
i = 2, j = 3, 4, 5, 6,\\
&i = 3, j = 2, 4, 5, 8,\quad
i = 4, j = 1, 4, 6, 7,\\
&i= 5, j = 1, 4, 6, 7,\quad
i = 6, j = 1, 3, 6, 8,\\
&i = 7, j = 1, 2, 7, 8,\quad
i = 8, j = 1, 2, 3, 4.
\end{aligned}
\tag{50}
\end{equation*}
Note that
\begin{equation*}
\begin{aligned}
&[\rho(\xi_3)^+, t^nE_1^{1,8}] = t^{n+1}(E_0^{1,2} + \tilde{E}_0^{3,8}),\\
&[t^{n}(E_0^{1,2} + \tilde{E}_0^{3,8}), {E}_1^{2,4}] = t^{n}E_1^{1,4},\\
&[\rho(\xi_2)^+, t^nE_1^{1,4}] = t^{n+1}\tilde{E}_0^{2,4}.
\end{aligned}
\tag{51}
\end{equation*}
On the other hand,
$$[\rho(\eta_1)^+, t^nE_1^{2,4}] = t^{n+1}({E}_0^{2,4} +  \tilde{E}_0^{2,4}).
\eqno (52)$$
Hence, $t^{n}{E}_0^{2,4}\in S$ and $t^n\tilde{E}_0^{2,4}\in S$.
Similarly, we can show that  all elements of  $E_0$ are in $S$.
According to (50), for each fixed $j$, where $1\leq j\leq 8$,
we have that $t^{n}{E}_1^{k,j}\in S$ for some $1\leq k\leq 8$.
Note also that for each fixed $j$, there exists $k$ such that
 $(t^{n}d){E}_1^{k,j}\in S$.
For example, the supercommutator
$$[t^n{E}_0^{5,2}, \rho(\xi_2)^+] = -(t^{n+2}d){E}_1^{5,1}\eqno (53)$$
produces such an element for $j = 1$.
Obviously, using  supercommutators of  $t^{n}{E}_0^{i,k}$ with
$t^n{E}_1^{k,j}$ and $(t^{n}d){E}_1^{k,j}$, we obtain that
$t^n{E}_1^{i,j}\in S$ and $(t^{n}d){E}_1^{i,j}\in S$ for any $i$.
Hence $E_1\subset S$. Finally, for each $\xi_i$ and
$ \eta_i$,
we now have that $\rho(\xi_i)_{-1}^+\in S$ and $\rho(\eta_i)_{-1}^+\in S$. From these matrices
we can obtain all matrices $t^{n}{E}_{-1}^{i,j}$
using supercommutators with
$t^{n}\tilde{E}_0^{i,j}$. Hence $E_{-1}\subset S$. Thus $S$ coincides with
$\hbox{End}(\W^{2^{N-1}|2^{N-1}})$.

\noindent
{\it Case $N > 4$}. Induction on $N$.
Assume that the statement is proved for $N-1$.
Let $C(2N-2)$ be the Clifford superalgebra with generators
$\xi_i$ and $\eta_i$, where $i = 1, \ldots, N-1$.
Present
$\Lambda(\xi_1, \ldots, \xi_{N})$ as follows:
\begin{equation*}
\begin{aligned}
&\Lambda(\xi_1, \ldots, \xi_{N})_{\bar{0}} =
\Lambda(\xi_1, \ldots, \xi_{N-1})_{\bar{1}} \xi_{N}\oplus
\Lambda(\xi_1, \ldots, \xi_{N-1})_{\bar{0}},\\
&\Lambda(\xi_1, \ldots, \xi_{N})_{\bar{1}} =
\Lambda(\xi_1, \ldots, \xi_{N-1})_{\bar{1}}\oplus
\Lambda(\xi_1, \ldots, \xi_{N-1})_{\bar{0}} \xi_{N}.
\end{aligned}
\tag{54}
\end{equation*}
Let
$$\hbox{End}(\W^{2^{N-1}|2^{N-1}}) =
\hbox{End}_{-1}\oplus \hbox{End}_0 \oplus \hbox{End}_1.\eqno (55)$$
Let ${E}_{\pm 1}^{i,j}$ ($1\leq i, j, \leq 2^{N-1}$)
 be an elementary $2^{N-1}\times 2^{N-1}$ matrix
in $\hbox{End}_{\pm 1}$, and let ${E}_{0}^{i,j}$ and
$\tilde{E}_{0}^{ij} = E_0^{i+2^{N-1},j+2^{N-1}}$ be elementary matrices in $\hbox{End}_0$.
By the inductive hypothesis,
$\s\p\o (2|2N-2)$ and $\tilde{\o}(2N-2, \C)$ generate
$$\hbox{End}(\W^{2^{N-2}|2^{N-2}}) =
\hbox{End}'_{-1}\oplus \hbox{End}'_0 \oplus \hbox{End}'_1,\eqno (56)$$
where
\begin{equation*}
\begin{aligned}
&\hbox{End}'_{-1} = \hbox{Span}(\W E_{-1}^{i,j+2^{N-2}})\subset \hbox{End}_{-1},\\
&\hbox{End}'_{1} = \hbox{Span}(\W E_{1}^{i+2^{N-2}, j})\subset \hbox{End}_1, \\
&\hbox{End}'_{0} = \hbox{Span}(\W E_{0}^{i+2^{N-2}, j+2^{N-2}},
\W\tilde{E}_{0}^{i,j}) \subset \hbox{End}_0,
\end{aligned}
\tag{57}
\end{equation*}
where $1\leq i, j \leq 2^{N-2}$.
Then  ${\o}(2N, \C)$ and $\hbox{End}(\W^{2^{N-2}|2^{N-2}})$
generate $\hbox{End}(\W^{2^{N-1}|2^{N-1}})$.
$$\eqno\Q$$

Note that we can realize
$\s\p\o(2|2N)$  as a subsuperalgebra of $K(2N)\subset P(2N)$.
$\s\p\o(2|2N)_{\bar{1}}$ is spanned by the elements
\begin{equation*}
\begin{aligned}
&(\xi_i)^{+}_K := t\xi_i + {1\over 2}\tau^{-1}(\sum_{j=1}^N\eta_j\xi_j)\xi_i,\quad
(\xi_i)^{-}_K := t^{-1}\xi_i - {1\over 2} t^{-2}\tau^{-1}(\sum_{j=1}^N\eta_j\xi_j)\xi_i,\\
&(\eta_i)^{+}_K := t^2\tau\eta_i + {1\over 2}t\eta_i(\sum_{j=1}^N\eta_j\xi_j) +
{1\over 2}\tau^{-1}\eta_i(\sum_{j,k=1}^N\eta_j\xi_j\eta_k\xi_k), \\
&(\eta_i)^{-}_K := \tau\eta_i - {1\over 2}t^{-1}\eta_i(\sum_{j=1}^N\eta_j\xi_j) +
{1\over 2}t^{-2}\tau^{-1}\eta_i(\sum_{j,k=1}^N\eta_j\xi_j\eta_k\xi_k),\\
\end{aligned}
\tag{58}
\end{equation*}
where $i = 1,\ldots, N$.
Let
\begin{equation*}
\begin{aligned}
\sigma: \s\p\o(2|2N)\longrightarrow\hbox{End}(\W^{2^{N-1}|2^{N-1}})
\end{aligned}
\tag{59}
\end{equation*}
be an embedding defined  by
\begin{equation*}
\begin{aligned}
\sigma((\xi_i)^{\pm}_K) = \rho (\xi_i)^{\pm}, \quad \sigma((\eta_i)^{\pm}_K) = \rho (\eta_i)^{\pm}, \quad i = 1, \ldots, N,\quad  N\geq 1.
\end{aligned}
\tag{60}
\end{equation*}
Note that embeddings of
$K(2)$, $\hat{K}'(4)$ and $CK_6$, obtained  in Theorem 3.2, are extensions of $\sigma$, see (29), (38) and (44).

\noindent
{\bf Corollary 3.3.}
The embedding (59)
cannot be extended to an embedding of $K(2N)$ into $\hbox{End}(\W^{2^{N-1}|2^{N-1}})$, if $N\geq 4$.

\noindent
{\bf Proof.}
Suppose that there exists an embedding
\begin{equation*}
\begin{aligned}
\sigma: K(2N) \rightarrow \hbox{End}(\W^{2^{N-1}|2^{N-1}}).
\end{aligned}
\tag{61}
\end{equation*}
Then $\sigma(\s\p\o(2|2N))$ and $\hbox{Span}(t^n\sigma({\o}(2N, \C)))$
must generate a subsuperalgebra of

\noindent
$\sigma(K(2N))$,
which is not possible, since they generate the entire $\hbox{End}(\W^{2^{N-1}|2^{N-1}})$.
$$\eqno\Q$$
{\bf Remark 3.4.} Note that certain exceptional simple finite-dimensional Lie superalgebras
can also be realized as  subsuperalgebras of matrices over $\W$.
In [19] we obtained a realization of the family $D(2, 1; \alpha)$ as
$4\times 4$ matrices over $\W$.
Recall that $F(4)$ is an exceptional
finite-dimensional Lie superalgebra such that
$$F(4)_{\bar{0}} = \o(7)\oplus \s\l (2), \quad F(4)_{\bar{1}}= spin_7\otimes \s\l(2),$$
see [7]. It can be constructed using Clifford algebra techniques, see [21].
We conjecture that $F(4)$ can be realized as a subsuperalgebra of
matrices of size $16\times 16$ over $\W$, so that
$F(4)_{\bar{1}}$ is in the span of the matrix fields
generated by
$\rho(\xi_i)^{\pm}$ and $\rho(\eta_i)^{\pm}$, $i = 1, \ldots, 4$, under the action of $\tilde{\o}(8, \C)$.

\vfil\eject
\noindent
{\bf Acknowledgments}
\vskip 0.2in

The author is grateful to the
Institute for Advanced Study for the hospitality and support during term II of the academic year 2006--2007.

This material is based upon work supported by the National Science Foundation under agreement
$No.\hbox{ } DMS-0111298$. Any opinions, findings and conclusions or recommendations expressed in this material are those of the author and do not necessarily reflect the views of the National Science Foundation.

\vskip 0.2in
\noindent
{\bf References}
\vskip 0.2in

\begin {itemize}

\font\red=cmbsy10
\def\~{\hbox{\red\char'0016}}

\item[{[1]}]
M. Ademollo, L. Brink, A. D'Adda, R. D'Auria, E. Napolitano, S. Sciuto, E. Del Giudice, P. Di Vecchia, S. Ferrara, F. Gliozzi, R. Musto and R. Pettorino,
Supersymmetric strings and colour confinement,
Phys. Lett. B 62 (1976) 105--110.

\item[{[2]}]
M. Ademollo, L. Brink, A. D'Adda, R. D'Auria, E. Napolitano, S. Sciuto, E. Del Giudice, P. Di Vecchia, S. Ferrara, F. Gliozzi, R. Musto, R. Pettorino and J. Schwarz,
Dual strings with $U(1)$ colour symmetry,
Nucl. Phys. B 111 (1976) 77--110.

\item[{[3]}]
S.-J. Cheng and V. G.  Kac,
A new $N = 6$ superconformal algebra,
Commun. Math. Phys. 186 (1997)  219--231.

\item[{[4]}] B. Feigin and D. Leites,
New Lie superalgebras of string theories,
in:  Group-Theoretical Methods in
Physics, edited by  M. A. Markov, V. I. Man'ko and A. E. Shabad.
(Nauka, Moscow, 1983), Vol. 1, 269--273.
[English translation  Gordon and Breach, New York, 1984].

\item[{[5]}] M. Goto, F. Grosshans, Semisimple Lie algebras,
Lecture Notes in Pure and Applied Mathematics, Vol. 38.
Marcel Dekker, Inc., New York-Basel, 1978.

\item[{[6]}] P. Grozman, D. Leites, and I. Shchepochkina,
Lie superalgebras of string theories,
Acta Math. Vietnam. 26 (2001)  27--63; e-print arXiv:hep-th/9702120.

\item[{[7]}]
V. G. Kac, Lie superalgebras,
Adv. Math. 26  (1977) 8--96.

\item[{[8]}]
V. G. Kac,
Vertex Algebras for Beginners,
University Lecture Series, Vol. 10, American Mathematical Society, Providence, RI, 1996.
(Second edition, 1998).

\item[{[9]}]
V. G. Kac,  Superconformal algebras and transitive group actions on quadrics,
Commun. Math. Phys.  186 (1997) 233--252. Erratum  Commun. Math. Phys. 217 (2001) 697-698.

\item[{[10]}]
V. G. Kac,
Classification of infinite-dimensional simple linearly
compact Lie superalgebras,
Adv. Math. 139 (1998) 1--55.

\item[{[11]}]
V. G. Kac,
Structure of some $\Z$-graded Lie superalgebras of vector fields,
Transform. Groups  4 (1999) 219--272.

\item[{[12]}]
V. G. Kac,
Classification of supersymmetries,
Proceedings of the International Congress of Mathematicians, Beijing, 2002 (Higher Education Press, Beijing, 2002), Vol. I, pp. 319--344.

\item[{[13]}]
V. G. Kac and J. W. van de Leur,
On classification of superconformal algebras,
in: Strings-88, edited by S. J. Gates, C. R. Preitschopf and W. Siegel.
(World Scientific, Singapore, 1989),  77--106.

\item[{[14]}]
B. Khesin, V. Lyubashenko, and C. Roger,
Extensions and contractions of the Lie algebra of
q-pseudodifferential symbols on the circle,
J. Funct. Anal. 143 (1997)  55--97.

\item[{[15]}] C. Martinez and E. I. Zelmanov,
Simple finite-dimensional Jordan superalgebras of prime characteristic,
J. Algebra 236 (2001) 575--629.

\item[{[16]}] C. Martinez and E. I. Zelmanov,
Lie superalgebras graded by $P(n)$ and $Q(n)$,
Proc. Natl. Acad. Sci. U.S.A  100 (2003)  8130--8137.

\item[{[17]}] E. Poletaeva,
A spinor-like representation of the contact
superconformal algebra  $K'(4)$,
J. Math. Phys.  42 (2001) 526--540; e-print arXiv:hep-th/0011100
and references therein.

\item[{[18]}] E. Poletaeva,
On the exceptional $N = 6$
superconformal algebra,
J. Math. Phys.  46 (2005) 103504, 13 pp.
Publisher's note, J. Math. Phys. 47 (2006) 019901;
e-print arXiv:hep-th/0311247.

\item[{[19]}] E. Poletaeva,
Embedding of the Lie superalgebra $D(2, 1 ; \alpha)$
into the Lie superalgebra of pseudodifferential symbols on $S^{1|2}$,
J. Math. Phys. 48 (2007) 103504, 17 pp.; e-print arXiv:0709.0083.

\item[{[20]}] E. Poletaeva,
On matrix realizations of the contact superconformal algebra
$\hat{K}'(4)$ and the exceptional $N = 6$ superconformal algebra,
DCDIS A Supplement, Advances in Dynamical Systems,
Vol. 14 (S2) 285--289, 2007;
\hfil\break
arXiv:0707.3097.

\item[{[21]}] M. Scheunert, W. Nahm, and V. Rittenberg,
Classification of all simple graded Lie algebras whose Lie algebra
is reductive. II. Construction of the exceptional algebras,
J. Math. Phys. 17 (1976) 1640--1644.

\item[{[22]}] I. Shchepochkina,
The five exceptional simple Lie superalgebras of vector fields,
e-print arXiv:hep-th/9702121.

\item[{[23]}] I. Shchepochkina,
The five exceptional simple Lie superalgebras of vector fields,
Funktsional. Anal. i Prilozhen.  33 (1999)  59--72.
[Funct. Anal. Appl.  33 (1999) 208--219].

\item[{[24]}] I. Shchepochkina,
The five exceptional simple Lie superalgebras of vector fields
and their fourteen regradings,
Represent. Theory  3 (1999) 373--415.

\end{itemize}

\end{document}